# Bridging the Cognitive Gap: Co-Designing and Evaluating a Voice-Enabled Community Chatbot for Older Adults

Feng Chen[1], Luna Xingyu Li[1], Ray-Yuan Chung[1], Wenyu Zeng[1], Yein Jeon[1], Yizhou Hu[2], Oleg Zaslavsky[3]

[1]Department of Biomedical Informatics and Health Education, University of Washington, Seattle, WA; [2]Elgin Park Secondary School, Surrey, BC, Canada; [3]Biobehavioral Nursing and Health Informatics Department, University of Washington, WA

**Abstract**

*Digital portals in retirement communities often create physical and cognitive barriers for older adults, leading to digital avoidance. Generative AI offers a solution by enabling natural language interaction, yet its adoption is hindered by the opaque, "Black Box" nature of these systems and lingering usability challenges. To address this, we evaluated a voice-enabled Large Language Model (LLM) chatbot at a continuing care retirement community in the Pacific Northwest. Through a mixed-methods Co-Design and Literacy Workshop (N=25), we applied a "Glass Box" approach combining multimodal accessibility with intentional AI education. The intervention significantly improved participants' technical understanding (p=0.004) and perceived transparency (p=0.001), shifting their interaction model from blind trust to informed reliance prioritizing verifiable evidence. While voice input reduced cognitive load, usability scores dropped significantly for users aged 80 and older (r=-0.50), indicating that truly age-inclusive AI must evolve beyond touch-based interfaces toward zero-touch navigation.*

**Introduction**

As the global population ages, the ability to "age in place" increasingly depends on navigating digital ecosystems [1,2]. In independent living communities, daily services, like dining reservations, maintenance requests, and event scheduling, have migrated from paper-based logs to comprehensive digital portals [3]. While these systems improve operational efficiency for administrators, they often function as barriers for the older elderly (adults aged 80+). Research on the digital determinants of health indicates that legacy information systems, often characterized by hierarchical menus, small touch targets, and rigid syntax, create significant technostress for older adults [4]. This can lead to digital avoidance, where residents retreat from using available community resources, inadvertently increasing their social isolation and dependency on staff [5]. Consequently, the design of these community technologies is not merely a matter of convenience but a critical factor in preserving autonomy and well-being.

Generative Artificial Intelligence (GenAI) and Large Language Models (LLMs) offer a transformative potential to dismantle these barriers. By shifting the interaction paradigm from command-based (requiring users to locate and click specific buttons) to intent-based (allowing users to state their goals in natural language), LLMs can theoretically bypass the usability flaws of legacy portals [6]. Recent innovations in healthcare have begun to leverage this potential; for example, Talk2Care demonstrated that LLM-based voice assistants could successfully facilitate communication between older adults and healthcare providers by reducing the cognitive load of data entry [7]. However, while such systems address the input problem, they do not fully resolve the trust problem. The adoption of AI assistants in non-clinical, daily living contexts remains low, often stalled by two distinct but interrelated challenges: physical usability barriers and the opacity of the "Black Box".

The first obstacle is physical. Most LLM interfaces still rely on text-based chat, which demands fine motor control for typing and visual acuity for reading text, faculties that notably decline in the 80+ demographic [8]. While voice assistants (VAs) have been proposed as a solution, their adoption has been inconsistent. Recent studies suggest that while older adults appreciate the hands-free nature of VAs, they often struggle with the wake word mechanics and lack the confidence to navigate errors when the system misunderstands them [6,7].

The second, and perhaps more nuanced, obstacle is the "Black Box" nature of AI. Older adults are often "digitally discerning", possessing a lifetime of experience that prioritizes verification and authoritative sources [9,10]. The stochastic nature of LLMs, where answers are generated without a clear source, conflicts with this demographic's mental model of reliable information [10]. The phenomenon of hallucinations, where models generate plausible but factually incorrect information, poses a severe risk to trust, especially in health or community contexts where accuracy is paramount [11]. Current informatics literature often treats usability (can I use it?) and transparency (do I understand how it works?) as separate domains. There is limited empirical evidence on how demystifying the underlying technology, explaining the logic of the Black Box, affects an older adult's willingness to adopt it.

To address these gaps, this study presents a mixed-methods evaluation of a voice-enabled LLM prototype designed for a continuing care retirement community in the Pacific Northwest. The site offers tiered care (independent living to skilled nursing) for adults aged 60-104. The predominantly Caucasian and Asian population frequently manages chronic conditions (e.g., arthritis, dementia). Unlike clinical assistants that focus on health monitoring, our system focuses on community information retrieval, a centralized portal for daily activities, event schedules, and dining menus, which are essential for independent living and social engagements. We posit that to build sustainable trust we must move from a "Black Box" design to a "Glass Box" approach, whose reasoning is transparent and interpretable to users [12], combining accessible voice interaction with educational transparency. This paper details the design of our chatbot system and analyzes the results of a Co-Design and Literacy Workshop (N=25). We report on how this multimodal intervention influenced residents' technical understanding, perceived transparency, and trust calibration, offering a roadmap for designing age-inclusive AI that is both usable and intelligible.

**Method**
Figure 1 summarizes our workshop-centered evaluation. Input from stakeholders, specifically, interviews with two community board members and three current residents prior to designing the workshop, guided an instruction-based LLM prototype and the Streamlit chatbot design. We then conducted a 90-minute in-person workshop. The session was attended by 35 residents and included an introduction, feature demonstration, and task-based use, accompanied by pre/post surveys and observation of task completion. Quantitative summaries and thematic analysis of open-ended feedback were used to synthesize key findings and identify priorities for the next iteration.

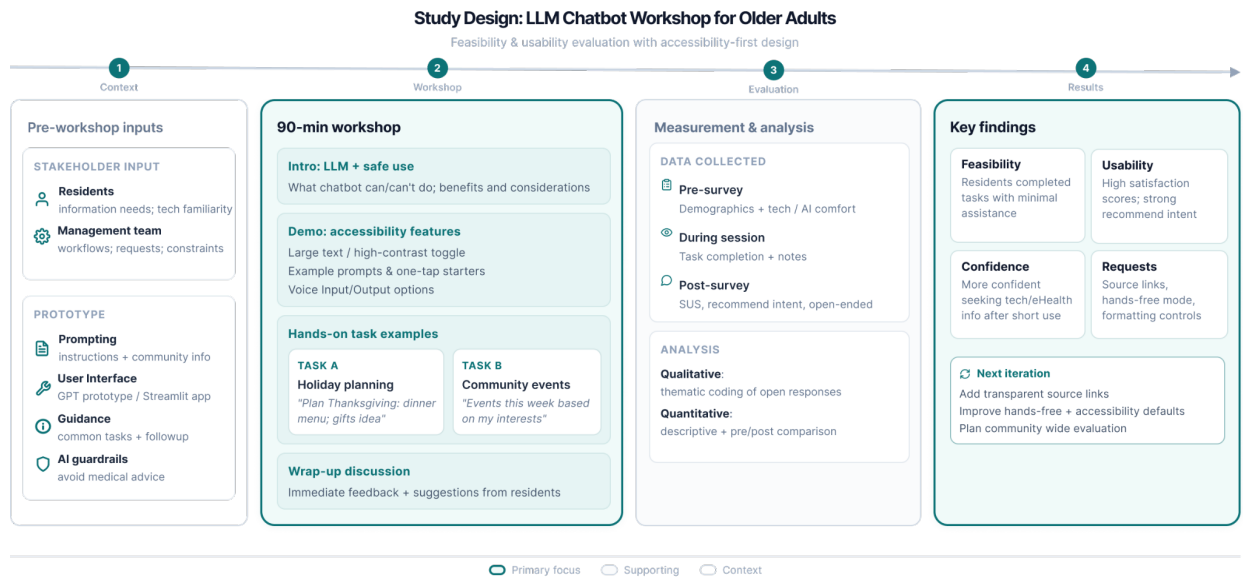

**Figure 1.** Study workflow for the LLM chatbot workshop. Pre-workshop stakeholder input informed an instruction-based LLM prototype and an accessibility-first Streamlit chatbot. A 90-minute hands-on workshop was evaluated using pre/post surveys and in-session task observations, followed by descriptive and thematic analysis to summarize feasibility/usability and guide iteration.

**Study Setting and Participants**
Participants were recruited from the independent living units via newsletters and digital signage 40 days prior to the workshop. Of the 35 residents who attended, 25 completed the surveys. The inclusion criteria required participants to be current residents of the independent living units and capable of providing informed consent. The resulting cohort was primarily composed of a highly educated group and skewed significantly toward an older population, with over half of the participants aged 80 and above. This specific demographic profile provided a unique opportunity to evaluate AI adoption among individuals who possess high cognitive literacy but face potential age-related sensory or motor declines. Table 1 summarizes the demographic characteristics of the participants.

**Table 1.** Participant Demographics (N=25) *One participant did not report age.

| Demographic | Category | N |
|---|---|---|
| Sex at Birth | Female | 16 |
| | Male | 9 |
| Age Group* | 70–79 | 10 |
| | 80 or older | 14 |
| Education Level | High School | 1 |
| | Bachelor | 6 |
| | Master | 8 |
| | Doctoral | 10 |

**System Design: The Voice-Enabled LLM-based Prototype**
To address the legacy portal's limitations, we developed an LLM-based chatbot through an iterative, user-centered design process. Prior qualitative interviews and needs assessments in this community demonstrated that legacy systems created significant information gaps and technostress, while conversational AI showed potential to improve technology literacy [13,14]. These foundational studies established the need for accessible gerontechnology, emphasizing prompt engineering to create a patient, helpful persona[15].

Building upon this foundational work, the current system utilizes GPT-4.1-mini as the backend model and introduces two critical technical advancements to address previously unresolved usability barriers. First, rather than relying on complex vector databases, the backend utilizes a Time-Aware Context Injection approach strictly indexed on the community's official event calendar. By restricting the LLM context to this curated source of truth and dynamically injecting the current date and time into the system prompt, the system minimizes hallucinations and accurately resolves relative temporal queries (e.g., "What community events do we have for today?"). Second, responding to feedback regarding the difficulty of typing on touchscreens and reading long responses, we integrated a multimodal interface. The prototype features a dedicated button for voice input utilizing Voice-to-Text (V2T) transcription, allowing residents to query the system using natural language. To further support accessibility, a "readout" button was included for each generated response, providing voice output via Text-to-Speech (Figure 2). This dual-voice capability allows residents to interact with community information without the need to type, navigate complex menus, or strain to read long responses. Finally, the system prompts enforced plain-language, stepwise responses, transparency about limitations, and domain scope restrictions. It was explicitly instructed to avoid providing advice on medications, diagnosis, or treatment; in such cases, it provided a brief limitation statement and suggested seeking appropriate professional support.

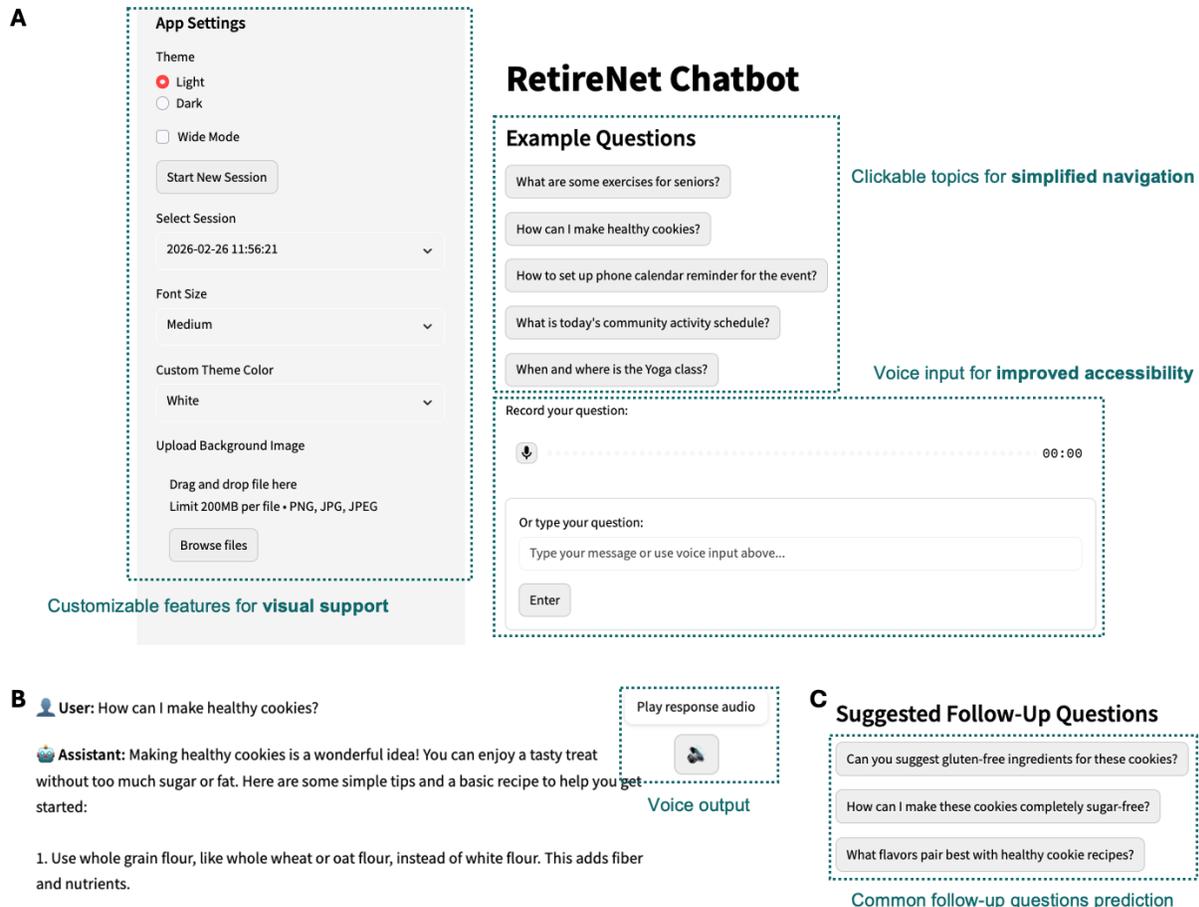

**Figure 2.** Screenshot of the implemented chatbot interface. (A) Startup page with setting options. (B) Chatbot response to an example question with audio support. (C) Clickable follow-up questions after each response.

**The Co-design and Literacy Workshop**
The intervention was operationalized as a single-session, 90-minute workshop designed to simultaneously educate residents on Generative AI concepts and evaluate the prototype's ecological validity. Prior to the start of the session, participants completed a pre-workshop survey to establish baseline metrics. The session then followed a structured three-phase protocol. The first phase, LLM introduction and Q&A (30 minutes), provided an educational overview of LLMs, explaining concepts such as "what is LLMs" and "What LLMs can/can't do" to remove the mystique often associated with AI. Facilitators explicitly discussed the limitations of LLMs, including the potential for error, to frame residents' expectations realistically. The second phase, Prototype Demonstration (15 minutes), showcased our chatbot's capabilities, specifically highlighting the new Voice-to-Text feature and accessibility settings such as dynamic font scaling. The final phase, Hands-On Exploration and Q&A (45 minutes), allowed participants to engage in free exploration using their own devices or provided tablets. Residents were encouraged to complete ecological tasks relevant to their daily lives, such as finding a holiday recipe or checking a community schedule, rather than scripted test cases, ensuring the feedback reflected real-world usage patterns. Immediately following this exploration phase, the post-workshop survey was distributed to capture immediate feedback and usability metrics.

**Data Collection Instruments**
This study was reviewed and determined to meet exempt status by the University of Washington Institutional Review Board (IRB #STUDY00024237). Participants' names and contact information were collected for workshop communication and follow-up purposes. Access to identifiable information was restricted to authorized study personnel. A mixed-methods approach captured usability metrics and user sentiment. The pre-workshop survey collected demographic data, AI familiarity, and established baselines for technology usage and AI trustworthiness using paired 5-point Likert-scale questions. The post-workshop survey was designed to assess changes in attitude using the same set of paired questions. Crucially, this post-session instrument also included the standardized System

Usability Scale (SUS) [16], a 10-item questionnaire yielding a composite score from 0 to 100, as well as open-ended questions soliciting qualitative feedback regarding user experience and design frustrations.

**Data analysis**
All survey responses and qualitative feedback were manually coded into Microsoft Excel. To ensure confidentiality, all data were de-identified using unique participant identifiers. Quantitative statistical analysis was performed using Python (v3.9.19) with the pandas (v2.2.3) and scipy (v1.12.0) libraries. Descriptive statistics (mean, standard deviation, and frequency distributions) were calculated to summarize participant demographics and baseline technology usage. To evaluate the impact of the workshop, we conducted paired sample t-tests comparing pre- and post-workshop responses for attitude and confidence metrics. Effect sizes were determined using Cohen's d to interpret the magnitude of observed changes. We also assessed the relationship between age, prior AI knowledge, and SUS using Pearson correlation coefficients (r). A p-value of less than 0.05 was considered statistically significant for all tests. Qualitative feedback from open-ended survey questions was analyzed using an inductive thematic analysis approach [17]. Responses were first reviewed for familiarity, then explicitly coded to identify recurring sentiments regarding specific features (e.g., voice input) and conceptual concerns (e.g., data privacy). These codes were subsequently grouped into broader themes to contextualize the quantitative findings, specifically focusing on the disconnect between usability and trust.

**Result**
**Participant Characteristics and Baseline Technology Profile**
The cohort (N=25) demonstrated high baseline digital literacy but varied AI experience (Figure 3). While internet usage was nearly universal (84% multiple times daily) and 56% currently used voice assistants, participants entered the workshop with significant humility regarding AI expertise; 92% described their knowledge of AI as limited or nonexistent. Only one participant reported being very familiar with the technology, and no participants identified as experts. This combination of high internet literacy and low AI-specific knowledge provided a fertile ground for evaluating how a structured educational intervention might alter perceptions of trust and usability.

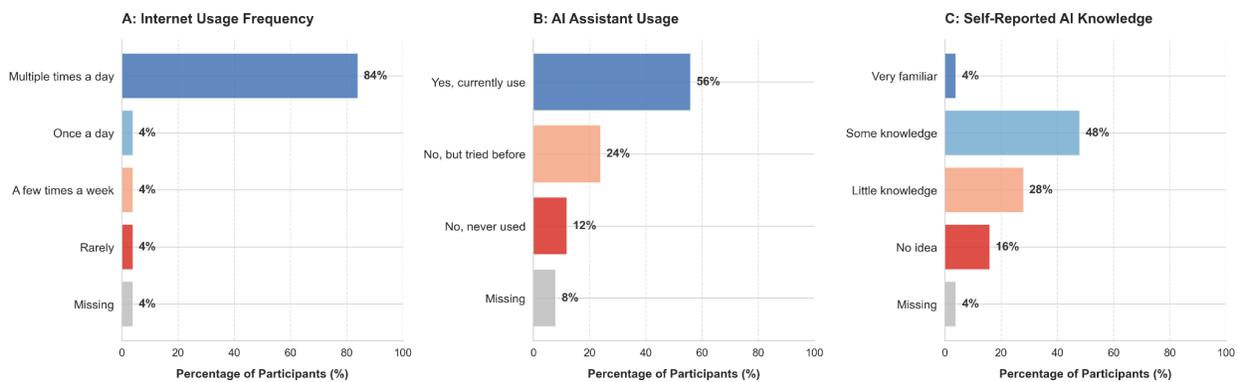

**Figure 3.** Baseline technology profile of participants (N=25). (A) Daily internet usage frequency, indicating high general digital engagement. (B) Current adoption rates of AI voice assistants. (C) Self-reported knowledge of AI and chatbots, demonstrating a significant gap between general internet literacy and AI-specific expertise.

**Quantitative Findings**
The System Usability Scale (SUS) analysis yielded a mean score of 60.54 (SD=18.37). According to standard SUS benchmarks, this score indicates marginal usability, falling below the industry average of 68. However, a significant negative correlation (r = -0.5) was observed between age and SUS scores, revealing a usability cliff. Participants in the 70–79 age group reported higher satisfaction, while residents aged 80 and older experienced markedly higher friction. Interestingly, prior AI knowledge showed a negligible correlation (r = -0.12, p>0.05) with usability, suggesting that physical interface challenges outweighed any conceptual familiarity with the technology.

Paired t-tests comparing pre- and post-workshop responses revealed that the intervention was most effective at increasing transparency and understanding rather than blind trust, as shown in Figure 4. The most significant improvements were observed in participants' understanding of why the chatbot gave certain answers (+0.93 average increase, p=0.004, Cohen's d=0.93) and the perception that the chatbot was transparent about its logic (+0.87 average increase, p=0.001, Cohen's d=1.04). Confidence in using AI systems also saw a statistically significant increase (+0.63 average increase, p=0.009, Cohen's d=0.66). However, general "Trust in AI" showed a smaller, non-significant increase (p=0.11). This suggests that while the workshop successfully demystified the chatbot's mechanics and increased self-efficacy, it did not immediately overcome the participants' fundamental caution regarding AI reliability. By better understanding how the system generates answers, residents may have developed a more critical perspective, recognizing the tool's utility while still requiring consistent accuracy or verifiable sources before showing full trust towards the LLM-based chatbot.

The study also identified notable differences in attitude shifts based on sex. Male participants showed a higher mean improvement in their overall attitude toward the AI system (+0.83) compared to female participants (+0.23). Specifically, the male cohort in this study appeared more receptive to shifting their baseline attitudes after hands-on exploration, whereas the female cohort maintained a more stable and cautious perception. Future research with a larger sample size could further investigate whether these trends persist and if gender-specific pedagogical approaches are necessary to improve AI adoption in retirement communities.

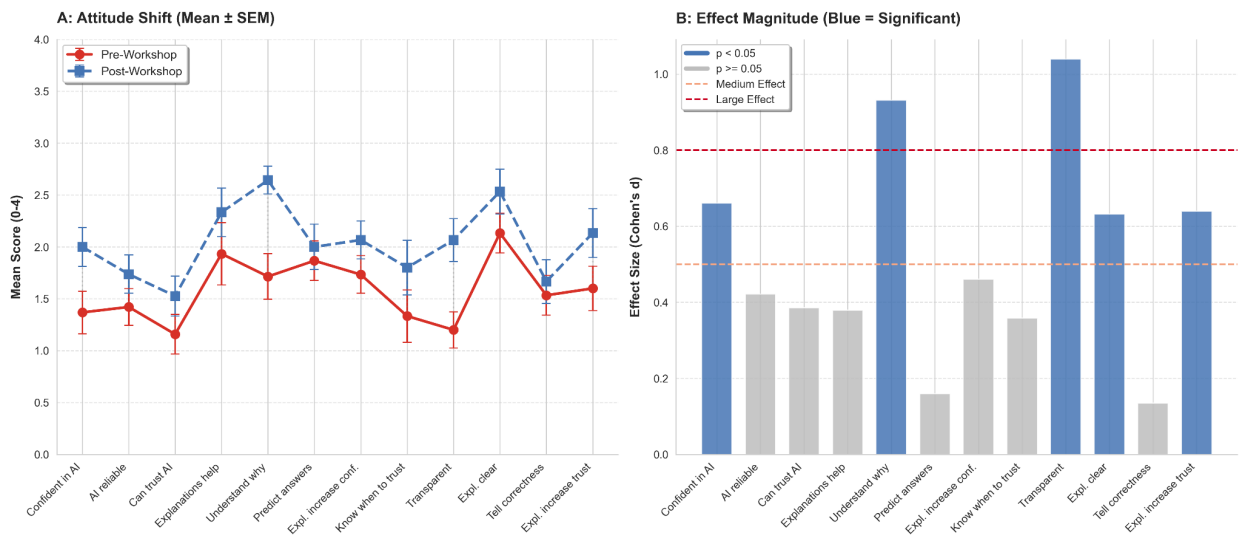

**Figure 4.** Comparison of pre- and post-workshop survey results assessing participant attitudes toward the AI chatbot.

**Qualitative Findings**
The thematic analysis of the open-ended survey responses identified four primary areas of focus for the residents, as shown in Table 2: the role of the chatbot as a bridge for existing digital systems, the requirement for verifiable evidence, the challenge of interface friction, and the boundaries of safe information categories.

Participants discussed the chatbot's value in relation to the community's primary digital portal. Many residents described the existing system as outdated or difficult to navigate, framing the chatbot as a necessary tool to simplify information retrieval. For these users, the primary appeal was the ability to bypass complicated menus to find specific facts like event schedules or community logistics. This suggests that the chatbot serves as a functional layer that improves accessibility for less intuitive legacy software.

Regarding the nature of trust, the qualitative data indicates that residents approach AI with significant discernment rather than blanket acceptance. Several participants emphasized that they would not rely on the system's output without the ability to verify it through original sources. The requirement for explicit citations was a recurring sentiment, as residents expressed a strong preference for seeing where the information originated to ensure accuracy. This suggests that while the workshop improved their understanding of how the system works, it also reinforced their preference for a verification-based model where the AI provides evidence to support its claims.

The introduction of voice interaction was viewed as a beneficial feature, yet it also highlighted persistent interface friction. While residents noted that speaking questions was an efficient and helpful option, they also reported frustrations with navigating the chatbot UI, such as "having to scroll through chatbot" or the precision required for

"the double-clicking button". These difficulties suggest that multimodal design must address more than just the initial query. If the subsequent navigation still requires fine motor control that is difficult for older adults, the overall benefit of the voice input and output is significantly reduced.

Finally, residents clearly distinguished between different categories of information based on perceived risk. While they were comfortable using the chatbot for logistical tasks like finding community activities or repair instructions, they were vocal about avoiding the system for personal, medical, or financial matters. This boundary indicates a sophisticated risk assessment by the participants. They view the chatbot as a convenient assistant for community navigation but maintain a firm requirement for human expertise and more secure channels when dealing with sensitive or life-critical information.

**Table 2.** Qualitative Themes, Participant Sentiments, and Design Recommendations for the Community Chatbot.

| Theme | Participant Sentiment | Recommendation for Design |
| --- | --- | --- |
| Utility as a Digital Bridge | "The current platform is terrible, this is better." | Position the chatbot as a smart search tool and functional layer over the existing community portal. |
| Interface Friction | "Double-clicking is hard; scrolling is tedious." | Simplify digital triggers; implement auto-listening and auto-scrolling to minimize fine motor requirements. |
| Demand for Verification | "Show me the sources." | Integrate direct hyperlinks to the community handbook or external authoritative sources within responses. |
| Safety Guardrails | "Not for medical or financial use." | Add clear disclaimers and warnings when users ask "High-Stakes" questions. |

**Discussion**

The primary objective of this study was to evaluate whether a "Glass Box" approach, combining accessible multimodal design with intentional education, could shift perceptions of AI among older adults. The results suggest that while a short-term intervention may not immediately generate deep trust, it is highly effective at increasing technical understanding and perceived transparency. The statistically significant improvements in participants' ability to understand why the chatbot provided certain answers ($p=0.004$) and their perception of the system as transparent ($p=0.001$) indicate that older adults are capable of moving beyond a "Black Box" mental model when provided with contextually grounded explanations. This aligns with recent research in Explainable AI (XAI), which posits that for older users, transparency is not merely a technical requirement but a psychological prerequisite for agency and autonomy [18]. By demystifying the underlying mechanics of the chatbot, the workshop transformed the new technology from a distant fantasy into a logical tool that can support their daily life.

While the workshop successfully transitioned participants from a "Black Box" to a "Glass Box" model, the lack of a significant increase in general trust ($p=0.11$) suggests that technical transparency does not automatically translate into unconditional acceptance. This result illustrates how users adjust their trust as they gain a better understanding of the technology. Rather than relying on general expectations, participants began to match their level of trust to the system's actual performance [19,20]. By learning the logic behind the chatbot, residents were able to distinguish between what the system can realistically achieve and what it cannot, leading to a more grounded form of reliance.

V2T bridged a functional gap, as touchscreen typing was a significant deterrent. The praise for voice input and output aligns with previous research suggesting that natural language interfaces can reduce the cognitive and motor load required for digital interaction by allowing users to bypass the rigid syntax and precise motor control needed for traditional menus [21]. However, the mean System Usability Scale (SUS) score of 60.54 and the strong negative correlation between age and usability ($r = -0.50$) reveal that a gap in accessibility remains for the older participants. These findings suggest that while voice input simplifies the initial query, the overall experience is still hindered by interface friction. Qualitative reports of frustration with scrolling through long text responses or the precision required for button interactions indicate that further improvement of the accessibility for the chatbot is still needed. For residents aged 80 and older, a truly accessible system may need to move toward a model where both interaction and navigation are handled verbally, decoupling the experience entirely from traditional touch-based requirements[22].

These results also offer a roadmap for the future of informatics in independent living environments. Notably, the data revealed a contrast between participants' nearly universal daily internet usage and their low baseline adoption of AI chatbots, despite being a highly educated cohort. This indicates that while older adults are highly accustomed to conventional information retrieval (e.g., standard search engines), natural language AI interaction remains a novel paradigm that they approach with caution. As evidenced by the clear boundaries residents set, welcoming help with community logistics while rejecting AI for medical or financial matters, this hesitation is likely rooted in a sophisticated risk assessment rather than a lack of digital literacy. Apprehensions regarding data privacy, the opacity of AI companies, and the security of personal information likely drive this reluctance to adopt conversational agents. This suggests that the most immediate, high-impact application for GenAI in this sector is not necessarily social companionship, but the semantic wrapping of poorly designed community data to make it accessible through natural language. This supports a step-by-step deployment strategy where AI first establishes reliability in low-stakes information retrieval before moving into more sensitive domains.

Furthermore, while the in-person workshop successfully implemented a "Glass Box" approach to debunk AI myths and increase technical understanding, facilitating physical coaching sessions is not always feasible in broader settings. This raises an important design question for scalability: should future AI products integrate mandatory digital tutorials or onboarding modules before allowing older adults to interact with the system? While embedding these educational steps directly into the software could proactively address user concerns and establish baseline trust, designers must carefully balance this against the risk of adding cognitive burden. For a demographic that already experiences friction with complex digital interfaces, a lengthy prerequisite tutorial might inadvertently increase technostress and lead to early abandonment. Future research should explore how to seamlessly weave transparent explanations into the natural flow of the chatbot interface, providing proper education that builds informed reliance without overwhelming the user.

**Limitations**
This study has several limitations that should be acknowledged. The sample size (N=25) was small and drawn from a single community, which skews toward a highly educated demographic that may not represent the general aging population. Furthermore, while 35 residents attended the workshop, only 25 completed the survey. This attrition likely introduces a selection bias; the residents who opted out of the survey may represent those with the highest levels of technostress or apprehension toward AI. Consequently, our findings may not fully capture the perspectives of the most hesitant users, meaning the reported usability and trust scores could represent a best-case scenario for this community. Additionally, the single-session format measures immediate attitude shifts but cannot predict long-term retention or habit formation. While the results show a promising increase in confidence and understanding, a one-time workshop is only the first step toward digital fluency. Future longitudinal studies are necessary to determine if these gains persist once residents return to their daily routines without the scaffolding of a structured environment. Investigating whether increased transparency leads to sustained usage or if the physical barriers identified ultimately lead to long-term digital avoidance will be essential for the continued development of age-inclusive artificial intelligence.

**Conclusion**
This study demonstrates that a "Glass Box" approach, combining accessible design with intentional education, can effectively bridge the digital divide for older adults. The significant gains in technical understanding and perceived transparency indicate that demystifying AI and LLMs is a critical first step toward inclusion. While general trust did not increase significantly, the intervention shifted how participants engaged with the system. Instead of viewing the LLM-based chatbot as a mysterious "Black Box", residents began to approach it with a more critical and practical mindset, focusing more on the accuracy of the information provided. This research highlights the value of an LLM-based chatbot as a useful tool for retrieving retirement community information and giving daily support. The chatbot reduced the cognitive load required to navigate community logistics. However, the marginal usability scores and the friction reported by participants aged 80 and older suggest that voice input and output alone is not enough. Truly inclusive design must move toward zero-touch navigation to fully accommodate the motor and sensory declines of the older population. Ultimately, prioritizing transparency and verifiable evidence provides a roadmap for building AI that supports long-term autonomy and confidence in retirement communities.

## Code Availability
Our code for prototype development and prompt engineering was available: https://github.com/chenfeng1234567/llm_chatbot.